# Charge-order-enhanced capacitance in semiconductor moiré superlattices

Tingxin Li[1], Jiacheng Zhu[1], Yanhao Tang[1], Kenji Watanabe[2], Takashi Taniguchi[2], Veit Elser[3], Jie Shan[1,3,4*], Kin Fai Mak[1,3,4*]

[1] School of Applied and Engineering Physics, Cornell University, Ithaca, NY, USA
[2] National Institute for Materials Science, 1-1 Namiki, 305-0044 Tsukuba, Japan
[3] Laboratory of Atomic and Solid State Physics, Cornell University, Ithaca, NY, USA
[4] Kavli Institute at Cornell for Nanoscale Science, Ithaca, NY, USA
*Email: jie.shan@cornell.edu; kinfai.mak@cornell.edu

**Van der Waals moiré materials have emerged as a highly controllable platform to study the electronic correlation phenomena [1-17]. In particular, robust correlated insulating states have recently been discovered at both integer and fractional filling factors of the semiconductor moiré systems [10-17]. Here, we reveal the thermodynamic properties of these states by measuring the gate capacitance of $MoSe_2/WS_2$ moiré superlattices. We observe a series of incompressible states for filling factor 0 – 8 and anomalously large capacitance (nearly 60% above the device's geometrical capacitance) in the intervening compressible regions. The anomalously large capacitance is most pronounced at small filling factor, below the melting temperature of the charge-ordered states, and for small sample-gate separation. It is a manifestation of the device-geometry-dependent Coulomb interaction between electrons and phase mixing of the charge-ordered states. We have further extracted the thermodynamic gap of the correlated insulating states and the entropy of the capacitive device. The results not only establish capacitance as a powerful probe of the correlated states in semiconductor moiré systems, but also demonstrate control of the extended Coulomb interaction in these materials via sample-gate coupling.**

The capacitance of a parallel-plate capacitor is a thermodynamic probe of a sample that serves as one of the plate electrodes [18]. The electric potential $eV$ required to maintain a sheet charge carrier density $n$ in equilibrium is usually expressed as a sum of two independent contributions, $eV = \phi + \mu$ ($e$ denoting the elementary charge). Here $\phi = ne^2/C_g$ is the electrostatic potential from the ideal (or geometrical) capacitance per unit area $C_g = \varepsilon/d$ defined by permittivity $\varepsilon$ and thickness $d$ of the medium between the two electrodes, and $\mu$ is the chemical potential of the sample. The differential capacitance per unit area ($C = e\frac{dn}{dV}$) follows the relation $C^{-1} = C_g^{-1} + C_Q^{-1}$, where $C_Q = e^2 \frac{dn}{d\mu}$ is the sample's quantum capacitance that represents its electronic compressibility or thermodynamic density of states (DOS) [19]. For an incompressible (insulating) state, one anticipates $C < C_g$. The capacitance can also exceed $C_g$, which is frequently referred to as negative compressibility, as a result of electron Coulomb correlations in the sample [20-24]. Such a treatment is appropriate in the high-density regime ($nd^2 \gg 1$), where the interaction between electrons in the sample does not depend on device geometry [25]. For a moiré superlattice of period $a_M$ that is comparable to or larger than $d$, the system is in the low-density regime nearly for all filling factors of interest and the above treatment is no longer appropriate. In this work, we establish capacitance as a probe of the emergent electronic states in moiré superlattices in this new regime. We investigate the effect of screening of the extended Coulomb interaction by varying sample-gate separation $d$.

When two monolayers of semiconducting transition metal dichalcogenides (TMDs) are overlaid with a small twist angle or lattice mismatch, a moiré superlattice of long period is formed. The superlattice potential causes extra Bragg reflection of electrons and induces flat moiré minibands for enhanced correlations. Robust correlated insulating states have been observed at both integer and fractional filling factors of the superlattices [10-17]. A single-band extended Hubbard model with twofold valley pseudospin degeneracy is believed to capture the essential low-energy physics of TMD heterobilayers [26]. In particular, the insulating state at filling factor $\nu = 1$ (that is, one particle per moiré superlattice cell or half filling of the first moiré miniband) is understood to originate from a strong on-site Coulomb repulsion $U$ compared to the lattice kinetic hopping $t$ (Ref. 26, 27). The charge-ordered insulating states at fractional fillings highlight the importance of the extended Coulomb interaction in addition to $U$ (Ref. 10, 15, 28, 29). To date, detection of these states has heavily relied on optical techniques, largely due to challenges in achieving good TMD-metal contacts for electrical studies, particularly at low charge density and low temperature.

**Capacitance measurements**
Here we report capacitance of field-effect devices comprised of $MoSe_2/WS_2$ heterobilayers (Fig. 1a) as a function of filling factor spanning up to four electron moiré minibands ($\nu = 0 – 8$). Details on device fabrication and measurements are described in Methods. Angle-aligned $MoSe_2/WS_2$ heterobilayers form a triangular moiré superlattice with $a_M \approx 8$ nm (corresponding to a moiré density of $1.9 \times 10^{12}$ cm$^{-2}$) because there is ~ 4% lattice mismatch between these materials. Since $MoSe_2/WS_2$ has a type I band alignment [30], the lowest energy electron moiré minibands are formed by states near the conduction minimum of $MoSe_2$ and are weakly coupled to the states in $WS_2$ (Fig. 1b). We achieve good $n$-type contacts for capacitance measurements down to 10 K using platinum electrodes [31] and heavily doped contact region achieved under a large global back gate voltage $V_b$. The electron density in the heterobilayer is tuned by varying a local top gate voltage $V_t$. We measure the differential capacitance $C$ by detecting charge modulation on the top gate while applying a small AC voltage (10 mV in amplitude) to the heterobilayer. The charge modulation on the sample is about twice the value that is collected from the top gate since the sample is almost equally spaced from the top and back gates. We also monitor the penetration field capacitance by detecting the induced charge modulation on the top gate while applying an AC voltage to the back gate. The two results are fully consistent (Extended Data Fig. 1).

Figure 1d is the measured differential capacitance in the unit of $C_g$ for device 1 ($d/a_M \approx$ 1) as a function of top gate voltage at 10 K (see Extended Data Fig. 2 and 3 for basic characterization of the capacitance measurements). We observe a step increase of capacitance around – 4 V when the Fermi level enters the conduction band of $MoSe_2$. The capacitance plateaus out above ~ 0 V when the sample is heavily electron doped. We calibrate capacitance using these two limits: $C/C_g = 0$ when the Fermi level lies inside the superlattice band gap (> 1 eV) and the sample is insulating (incompressible); $C/C_g \approx$

1 when the sample is heavily doped and behaves as a good conductor. At intermediate gate voltages, we identify a series of capacitance dips (incompressible states). The most prominent ones appear equally spaced in gate voltage and are assigned integer fillings $v$ = 1, 2, 3 and 4 (Extended Data Fig. 4 for $v$ up to 8). The assignment is consistent with the known moiré density and the carrier density $n$ evaluated from the gate voltage and $C_g \approx 2.1 \times 10^{-7}$ Fcm$^{-2}$. The geometrical capacitance is determined from the permittivity and thickness of the gate dielectric (hexagonal boron nitride, $\varepsilon \approx 3\varepsilon_0$ with $\varepsilon_0$ denoting the vacuum permittivity). It was independently verified by including a reference capacitor on the measurement chip.

We determine the charge density of the sample as a function of gate voltage in Fig. 1e by integrating the differential capacitance of Fig. 1d. We calibrate the scale of $n$ by setting it to the moiré density at $v$ = 1. The incompressible states manifest a density plateau of different width in gate voltage, indicating that charging into the sample is suppressed. Incompressible states at commensurate filling factors including 1/3, 1/2, 2/3, 4/3 and 5/3 can be identified from capacitance. The states at $v$ < 1 are more robust than the states at $v$ > 1. These results are in good agreement with the reported charge-ordered insulating states in WSe$_2$/WS$_2$ moiré superlattices [10,15].

Remarkably, we observe anomalously large capacitance in the compressible regions between the incompressible states. The enhancement is particularly large at small doping densities with $C$ exceeding $C_g$ by ~ 30% for device 1 ($d/a_M \approx 1$). The enhancement gradually dies off with increasing density ($v$ > 2). In Fig. 2a we compare devices of different sample-gate separation ($d/a_M \approx$ 0.6, 0.8, 1.0 and 1.5). The capacitance enhancement increases with decreasing $d/a_M$. It is as high as 60% in device 4 ($d/a_M \approx$ 0.6) at 10 K. In addition, more incompressible states are discernable from capacitance at the same temperature in devices with larger $d/a_M$.

To illustrate the relationship between the observed capacitance enhancement and charge ordering, we examine the temperature dependence in Fig. 3a for device 1 ($d/a_M \approx 1$) (similar results for device 4 are shown in Extended Data Fig. 5). At 100 K only the $v = 1$ state is observed. With decreasing temperature, the incompressible states emerge one after another. Table 1 summarizes the transition temperature $T_C$ for the identified incompressible states. At the same time, the capacitance in the compressible regions (e.g. $v$ = 0.42) increases quickly below $T_C$ of the fractional states (also see Extended Data Fig. 5), demonstrating capacitance enhancement from charge order formation (Fig. 3b). A small capacitance enhancement remains above $T_C$ after these states are melted.

**Charge-order-enhanced capacitance**
The experimental results show that the correlation effects are strongly dependent on sample-gate separation in devices with $d/a_M$ ~ 1, particularly, for the fractional-filling states since the gate electrodes effectively screen the extended Coulomb interactions. In this regime, it is no longer appropriate to separate the electric potential into independent contributions of the electrostatic energy from the ideal capacitance and the sample chemical potential [25]. Instead, we need to describe the entire device as a system. The top gate voltage and capacitance of our dual-gate devices are related to the energy density $u$ of the entire device, $V_t = 2u'/e$ and $C^{-1} = 2u''/e^2$, where $u'$ and $u''$ are the first and second derivatives of $u$ with respect to electron density ($n$) in the sample (Methods). We model $u$ by treating electrons within the extended Hubbard model in the flat band limit, which has been successful in describing the TMD heterobilayer moiré systems in the strong interaction limit [15,17]. For $\nu < 1$, the ground states are charge-ordered states at commensurate filling factors and coexisting charge-ordered states at incommensurate filling factors. The energy density is a piecewise linear function of density, whose slopes are given in terms of the energies of the adjacent charge-ordered states that participate in the mixture (Methods). The density dependence of $V_t$ exhibits a series of plateaus connected by voltage jumps at the charge-ordered states. The capacitance is zero for the charge-ordered states and infinity in between.

The measured capacitance deviates significantly from this ideal limit. It is finite at the charge-ordered states and exceeds $C_g$ by no more than 60% in the compressible regions. Finite temperature effects alone cannot explain the discrepancy because the energy scale corresponding to the lowest measurement temperature (10 K) is much smaller than the nearest-neighbor Coulomb repulsion (~ 50 meV [15]). The quantum effects from lattice hopping ($t \sim$ 1 meV [27]) are also too small to account for it. We consider the effects of inhomogeneous strain or twist angle disorders that could be introduced unintentionally during fabrication [32] by including an on-site disorder energy that varies smoothly over the scale of the moiré period. By numerically solving for the coexistence of previously identified charge-ordered states [15] in the presence of disorder we determine a $u''$ that can be compared with experiment (Methods).

Figure 2b illustrates the effect of on-site energy disorder for $d/a_M$ = 1. Disorders broaden the charge-ordered states so that weaker states (e.g. $\nu = 1/4$) could be smeared out. Disorders also introduce metallic regions in the sample, which acquire an electrostatic energy from charging and make the capacitance finite in the compressible regions. We find that root mean square (RMS) disorder potential strength $\varepsilon_S \approx$ 12.7 meV (~ 8% of the total moiré potential depth [33-35]) best describes our experiment. Figure 2c shows the result for different values of $d/a_M$ at a fixed $\varepsilon_S$ of 12.7 meV. It is in qualitative agreement with experiment (Fig. 2a). Particularly, with increasing $d/a_M$ the capacitance enhancement in the compressible regions decreases and the number of discernable charge-ordered states increases.

**Chemical potentials at integer fillings**
Next we consider the integer-filling states. The even integer states ($\nu$ =2 and 4) are moiré miniband insulators [26,27]. They are of single-particle origin. The odd integer states ($\nu$ = 1 and 3) are Mott or charge-transfer insulators [26,27]. They are largely driven by the on-site Coulomb repulsion, which is local and much less sensitive to the sample-gate separation compared to the extended Coulomb interaction. We thus expect the separation of the sample chemical potential $\mu$ from the electrostatic potential of the device to be approximately valid. We evaluate the sample chemical potential jump at each integer-filling state for adding or removing one electron by the area of the capacitance dip, $\Delta \mu = e \int dV_t (1 - C/C_g)/2$ (Methods). By extrapolating $\Delta \mu$ to the zero-temperature limit (Extended Data Fig. 6), we obtain the thermodynamic gap of the state (Table 1 for device 1). We were able to reliably exact $\Delta\mu$ from capacitance only for device 1, which has sufficiently small resistance at integer fillings (Extended Data Fig. 3).

We illustrate the schematic energy diagram of the MoSe$_2$/WS$_2$ moiré superlattice in Fig. 1c based on the energies inferred from capacitance. Since the interaction effect weakens with increasing $\nu$, we expect the Hubbard bandwidths to increase with $\nu$. The first gap (58.5 meV, $\nu = 1$) is the Mott or charge-transfer gap of the first moiré miniband. The value is comparable to but smaller than the on-site Coulomb repulsion $U \sim 200$ meV estimated from the moiré potential amplitude from optical and scanning tunneling microscopy measurements [33-35] (Methods). The first moiré miniband gap (36.3 meV, $\nu = 2$) is smaller than the Mott gap. This indicates the influence of remote bands, that is, mixing of different moiré minibands by $U$ (which can also reduce the measured Mott gap). The third gap (18.1 meV, $\nu = 3$ gap) is smaller than the $\nu = 1$ gap, which suggests a smaller on-site Coulomb repulsion for the second moiré miniband than the first. This result is consistent with the larger second moiré miniband bandwidth [26,27], which dictates a larger electron Wannier orbital size and a smaller on-site Coulomb repulsion. The weakened interaction effect is also consistent with the observed $C/C_g \to 1$ at large $\nu$ (Fig. 1d), where the sample and the gate electrode are decoupled in this limit.

**Device entropy**
Finally, we extract the entropy per moiré unit cell ($S$) of the capacitive device from the temperature dependence data in Fig. 3. At finite temperatures, the energy density of the device $u$ has to be replaced by the free energy density $f$, which gives us the Maxwell relation $\left(\frac{\partial V_t}{\partial T}\right)_\nu = \frac{2}{e}\left(\frac{\partial}{\partial n}\right)_T \left(\frac{\partial f}{\partial T}\right)_\nu = -\frac{2}{e}\left(\frac{\partial S}{\partial \nu}\right)_T$. Figure 3c shows the filling factor versus $V_t$ at varying temperatures for $\nu = 0 - 1$. The temperature dependence of $V_t$ (offset to zero at 10 K) at two representative filling factors is shown in the inset. To extract $S$, we estimate $\left(\frac{\partial V_t}{\partial T}\right)_\nu$ at every filling factor by the average slope over two temperature windows (due to limited signal-to-noise ratio): one below the $T_C$ of the charge-ordered states (15 – 40 K) and the other above $T_C$ (40 – 80 K). We then integrate $\left(\frac{\partial V_t}{\partial T}\right)_\nu$ over $\nu$ to obtain the entropy $S$ in Fig. 3d (Methods).

The entropy vanishes at $\nu = 0$ and 1 and shows local minimum at $\nu = 1/3$, 1/2 and 2/3 below $T_C$, as expected from the electronic DOS of MoSe$_2$/WS$_2$ moiré superlattices (the metal gate DOS is negligible). It reaches a maximum value $S \sim k_B$ near $\nu = 1/2$ ($k_B$ is the Boltzmann constant) and increases with temperature at all filling factors. No local minimum at fractional fillings can be identified above $T_C$. The results suggest faster entropy drop with decreasing temperature for the charge-ordered states. Future studies with improved sensitivity for measuring $\left(\frac{\partial V_t}{\partial T}\right)_\nu$ (e.g. temperature modulation studies [36]) may enable heat capacitance measurements of the charge-ordering transition. In conclusion, our study establishes capacitance as a powerful thermodynamic probe of the correlated states in semiconductor moiré superlattices. It also illustrates the importance of sample-gate coupling and the device-geometry-dependent extended Coulomb interaction at fractional fillings.

## Methods

**Device fabrication.** Angle-aligned MoSe$_2$/WS$_2$ moiré superlattices were assembled by the layer-by-layer dry transfer method as detailed in Ref. 37. Hexagonal boron nitride (hBN) was used as gate dielectric (out-of-plane component of $\varepsilon \approx 3\varepsilon_0$), whose thickness ranges from ~ 5 to 15 nm in different devices. Few-layer graphite or TaSe$_2$ (a TMD metal) was used as gate electrodes. Amorphous quartz with pre-patterned Ti/Pt electrodes was used as substrate to reduce the parasitic capacitance. We have measured 4 devices with $d/a_M \approx$ 0.6, 0.8, 1 and 1.5. Small device area of ~ 10 μm$^2$ was used to reduce the amount of spatial inhomogeneities. The optical image of a typical device is shown in Extended Data Fig. 7. To achieve good electrical contacts to the TMD moiré superlattices for capacitance measurements at low doping densities ($< 10^{12}$ cm$^{-2}$) and low temperatures (down to ~ 10 K), we applied a large positive global back-gate voltage (exact values are device dependent) to heavily dope the contact region to reduce the contact resistance. The top gate voltage $V_t$ was varied to tune the carrier density in the moiré superlattices while the back-gate voltage was fixed.

**Capacitance measurement.** The measurements were performed in a close-cycle He-4 cryostat (Oxford TeslatronPT). A commercial high electron mobility transistor (HEMT, model FHX35X) was used as the first-stage amplifier to effectively reduce the parasitic capacitance from cabling [38-40]. The work point of the HEMT was set by a voltage $V_H$ and a current $I_H$ shown in the schematic circuit diagram (Fig. 1a). Both the top-gate differential capacitance $C$ and the penetration differential capacitance $\tilde{C}$ were measured. The top gate capacitance is reported in the main text and the penetration capacitance is summarized in Extended Data Fig. 1. For the top gate capacitance measurements, we applied an AC voltage (10 mV in amplitude) to the TMD moiré superlattice and collected the signal from the top gate through the HEMT. The induced AC source-drain voltage across the HEMT was measured with a lock-in amplifier. For the penetration capacitance measurements, the AC voltage is applied to the back gate and the HEMT was connected to the top gate, from which the induced charge was collected.

To exclude potential contributions from the contact resistance and/or sample resistance to the measured capacitance (especially at low doping densities and low temperatures), we studied both the in-phase (capacitive) and out-of-phase (resistive) component of the signal as a function of frequency of the AC excitation voltage and sample temperature (Extended Data Fig. 2 and 3). At relatively high temperatures (> 10 K), the out-of-phase component is in general negligible compared to the in-phase component from 137 Hz to 2237 Hz. Below 10 K, the out-of-phase component gains importance with increasing measurement frequency at low doping densities ($|\nu| < 1$) and the results were not included in the main text. Also, for device 2-4, the resistive component at $\nu = 1$ and $\nu = 2$ is important up to ~ 20 K and down to the lowest measured frequency due to the large contact/sample resistance. Capacitance measurements and energy gap calculations are thus unreliable for the $\nu = 1$ and $\nu = 2$ insulating states. On the other hand, the resistive component for device 1 is negligible down to ~ 10 K at all filling factors (Extended Data

Fig. 3). We therefore use data from device 1 to evaluate the thermodynamic gap in Table 1.

**Estimate of $U$.** We estimate the magnitude of the on-site Coulomb repulsion for the first electron moiré miniband in MoSe$_2$/WS$_2$ moiré superlattices as $U = \frac{e^2}{4\pi\varepsilon\xi}$. Here $\varepsilon \approx 4.5\varepsilon_0$ is the effective permittivity of the background (hBN) and $\xi$ is the size of the lowest-energy local Wannier orbital. We estimate $\xi = \sqrt{a_M}\left(\frac{\pi^4 m V_M}{\hbar^2}\right)^{-1/4} \propto \sqrt{a_M}$ by approximating the moiré potential as a harmonic potential around its minimum. Here $a_M$ ($\approx 8$ nm) is the moiré period, $m$ ($\approx 0.5 m_0$) is the conduction band mass of monolayer MoSe$_2$, and $V_M$ is the amplitude of the moiré potential (the potential depth is up to $6V_M$ in this definition). The constants $e$, $\varepsilon_0$, $m_0$ and $\hbar$ denote, respectively, the elementary charge, the vacuum permittivity, the free electron mass and the reduced Planck's constant. We obtain $U \approx 210$ meV for $V_M \approx 20$ meV. The latter is inferred from both optical and scanning tunneling microscopy measurements of similar TMD moiré superlattices [33-35].

**Lumped circuit model for the capacitance measurement.** We consider the geometry where the sample is equally spaced from the top and back gates with distance $d$. A lumped circuit model for the capacitance measurement is appropriate (Extended Data Fig. 8) when the interaction between electrons in the sample does not dependent on the device geometry and the electric potential on the gate can be separated into independent contributions of the electrostatic energy from the ideal capacitance and the sample chemical potential. We derive from the circuit the top gate capacitance

$$\frac{1}{C} \approx \frac{2}{C_Q} + \frac{1}{C_g}, \quad (1)$$

and for the penetration capacitance $\tilde{C}$

$$\frac{1}{\tilde{C}} \approx \frac{2}{C_g} + \frac{C_Q}{C_g^2}. \quad (2)$$

Here $C_Q$ denotes the sample quantum capacitance and $C_g$ is the geometrical capacitance between the sample and the top (or back) gate electrode. All the capacitances are expressed in capacitance per unit area. An incompressible state appears as a dip in the top gate capacitance, and as a peak in the penetration capacitance, as a function of gate voltage.

As discussed in the main text, this approximation is appropriate for integer-filling states. The chemical potential jump $\Delta\mu$ at an integer-filling state can be obtained from the gate dependence of the top gate capacitance as

$$\Delta\mu = \int d\mu = \frac{e}{2}\int dV_t \left(1 - \frac{C}{C_g}\right), \quad (3)$$

and the penetration capacitance as

$$\Delta \mu = \int d\mu = e \int dV_t \frac{\tilde{C}}{C_g}. \tag{4}$$

The range of integration includes the corresponding capacitance dip/peak in gate voltage. The value of $\Delta \mu$ is half of the dip area from the top gate capacitance measurement, and the full peak area from the penetration capacitance measurement.

**General top gate capacitance model.** When the interaction between the charges in the sample is strongly dependent on the device geometry, it is no longer appropriate to separate $eV$ into independent contributions and apply the lumped circuit model [25]. Here we derive the capacitance as a function of the energy density $u$ of the entire device. In general, there is electron density $-n_t$ on the top gate, $-n_b$ on the back gate, and electron density $n_t + n_b = n$ on the sample. Defining $\Delta n = (n_t - n_b)/2$, we express the gate densities as $-n_t = -n/2 - \Delta n$, $-n_b = -n/2 + \Delta n$ and write the energy density of the entire device as

$$u(n_t, n_b) = u_s(n) + \frac{(e\Delta n)^2}{C_g}, \tag{5}$$

where $u_s(n)$ is the energy density of the sample when it holds electron density $n$ and is surrounded by gates with equal, compensating densities $-n/2$. The separation of the energy into the two independent contributions in Eqn. (5) follows from superposition of the corresponding electric fields. The field produced by $en$ is symmetric about the sample plane while electric field produced by $\pm e\Delta n$ on the two gates is antisymmetric with respect to that plane, with the result that there is no cross-term in the electric energy density.

The electron densities $-n_t$ and $-n_b$ on the top and back gates are kept in equilibrium by voltages $-V_t$ and $-V_b$ on these gates, relative to the sample, provided

$$V_t = \frac{1}{e}\frac{\partial u}{\partial n_t}, \qquad V_b = \frac{1}{e}\frac{\partial u}{\partial n_b}. \tag{6}$$

Using (5) these equations are equivalent to

$$\frac{V_t + V_b}{2} = u_s'(n_t + n_b)/e \tag{7}$$

$$\frac{V_t - V_b}{2} = \frac{e(n_t - n_b)}{2C_g}, \tag{8}$$

where $u_s' = \frac{du_s}{dn}$. Solving (8) for $n_b$ and substituting into (7) we obtain

$$\frac{V_t + V_b}{2} = u'_s(2n_t - C_g(V_t - V_b)/e)/e. \tag{9}$$

In general, the density response with respect to arbitrary variations of $V_t$ and $V_b$ can be obtained from this relation. In our experiment we measure the top gate differential capacitance ($C = e\frac{dn_t}{dV_t}$) under the constraint $dV_t = dV_b$ and $dn_t = dn_b = dn/2$ because the same AC voltage $dV_t$ is applied between the sample and the two gates. Applying this to Eqn. (9) we obtain

$$\frac{1}{C} = 2u''_s/e^2. \tag{11}$$

Following a similar procedure, we can also derive the penetration capacitance $\tilde{C} = -e\frac{dn_t}{dV_b}$ (with $dV_t = 0$)

$$\tilde{C} = \frac{C_g}{2} - \frac{e^2}{4u''_s} = \frac{1}{2}(C_g - C). \tag{12}$$

**Microscopic model for $u_s(n)$.** Geometrically, we model the device as a thin sample of area $A$ between top and back conducting gates equally spaced by distance $d$. The region between the conductors is filled with a uniform dielectric medium with permittivity $\varepsilon$. The moiré superlattice constant is $a_M$ and the unit cell area is $A_M = \sqrt{3}a_M^2/2$.

We analyze the zero temperature limit of the classical Hamiltonian

$$H = \frac{1}{2}\sum_{i \neq j} V(r_{ij}, d)\, n_i\, n_j + \sum_i D_i\, n_i, \tag{13}$$

where $n_i \in \{0,1\}$ is the occupation of moiré site $i$, $D_i$ is the on-site disorder energy at that site, $r_{ij}$ is the distance between sites $i$ and $j$, and

$$V(r, d) = \left(\frac{e^2}{4\pi\varepsilon}\right) \sum_{k=-\infty}^{\infty} \frac{(-1)^k}{\sqrt{r^2 + (2kd)^2}}, \tag{14}$$

is the electrostatic energy in the symmetrical two-conductor geometry. The ratio $u_s = H/A$ is the sample energy density defined above. All of our energies, when not explicitly specified, are in the energy unit

$$\epsilon_a = \frac{e^2}{4\pi\varepsilon a_M} \approx 48 \text{ meV}. \tag{15}$$

The values of $D_i$ are distributed in the range $[-\delta, \delta]$ with distribution function

$$\rho(\epsilon) = D\,(\delta^2 - \epsilon^2)^2, \tag{16}$$

where the normalization constant $D$ is chosen so that the integral of $\rho(\epsilon)$ over the dimensionless energy $\epsilon$ is 1. For this disorder distribution, the parameter $\delta = 0.7$ corresponds to an RMS disorder strength of 12.7 meV.

In the weak disorder limit we may use the following simple model for the ground states of $H$. Assuming that $D_i$ varies smoothly on the moiré superlattice, the ground states are co-existing charge ordered states with domains selected such that the densest state occupies the sites with lowest $D_i$, followed by the next-densest charge ordered state on the next-lowest sites, etc. Let $\nu_1 > \nu_2 > \cdots > \nu_{m+1}$ be the fillings of up to $m + 1$ charge ordered states that participate when the average filling is $\nu$, and the associated site-energy boundaries are $-\delta = \epsilon_0 \leq \epsilon_1 \leq \cdots \leq \epsilon_m \leq \epsilon_{m+1} = \delta$. The constraint that all electrons are in a charge ordered state with the correct (decreasing) order of fillings is expressed as

$$\nu = n(\epsilon_1, \cdots, \epsilon_m) = \sum_{k=1}^{m+1} \nu_k \int_{\epsilon_{k-1}}^{\epsilon_k} \rho(\epsilon) \, d\epsilon . \tag{17}$$

The energy per moiré site, for the same boundary energies, is

$$u_m(\epsilon_1, \cdots, \epsilon_m) = \sum_{k=1}^{m+1} \nu_k \int_{\epsilon_{k-1}}^{\epsilon_k} (\mu_k + \epsilon) \, \rho(\epsilon) \, d\epsilon , \tag{18}$$

where

$$\mu_k = \frac{1}{2} \sum_{i \neq 0} V(r_{i0}, d) \, n(k)_i \tag{19}$$

is the electrostatic energy of an electron at site 0 in the $k^{\text{th}}$ charge-ordered state with occupations $n(k)$ for which $n(k)_0 = 1$. The expression for $\mu_k$ needs to be averaged over the site 0 when the electrons in the charge ordered state are not equivalent by symmetry (e.g. $\nu = 3/5$). The chemical potentials labeled by filling, $\mu_\nu$, are given in Extended Data Table 1 (in units of $\epsilon_a$) for the fillings we identified in our earlier experiment [15] with $d/a_M = 5$, but now for the smaller values of this ratio relevant for the current experiments. Thermodynamic stability in the zero disorder limit requires that the chemical potential $\mu$ of a pure phase with filling $\nu$ is lower than the energy of a mixture formed by the adjacent phases with fillings $\nu_-$ and $\nu_+$ and corresponding $\mu_-$ and $\mu_+$. This translates to the inequality

$$(\nu_+ - \nu_-)\nu \, \mu < (\nu_+ - \nu)\nu_- \, \mu_- + (\nu - \nu_-)\nu_+ \, \mu_+ , \tag{20}$$

which is satisfied at all the fillings and the values of $d/a_M$ in Extended Data Table 1.

The ground state energy per unit area $u_s(n)$, for average filling $\nu$, is defined as the solution of a constrained minimization problem:

$$A_M \, u_s(\nu/A_M) = \min_{\substack{-\delta \leq \epsilon_1 \leq \cdots \leq \epsilon_m \leq \delta \\ \nu = n(\epsilon_1, \cdots, \epsilon_m)}} u_m(\epsilon_1, \cdots, \epsilon_m) . \tag{21}$$

We evaluated $u_s$ for our model as a function of $d/a_M$ for fillings in the range $0 < \nu < 1$ and for several strengths of disorder $\delta$. In the clean limit $\delta = 0$, $u_s$ is a piecewise linear function of density, given by weighting the pure-phase energies of the adjacent charge-ordered states. As discussed in the main text, the capacitance is zero for pure phases and infinity for mixtures.

With finite disorder $\delta > 0$, the constrained minimization Eqn. (21) was solved numerically on a fine grid of $\nu$ values with the `FindMinimum` function of *Mathematica*, and the second derivative $u_s''$ was computed by finite differences. Selected results are shown in Fig. 2. In addition to the incompressible $\nu = 1$ state, we also see incompressible behavior at $\nu = 0, 1/3, 2/3$, and to a smaller degree, $\nu = 1/2$. Incompressibility at $\nu = 0$ is strictly a consequence of disorder. The pockets of $\nu = 1/7$ domains in the $\nu \to 0$ limit of our model are just individual electrons optimally occupying the sites with lowest disorder potential. If disorder played no role in the experiment, then the resulting incommensurate Wigner crystal for $\nu \to 0$ would exhibit a diverging compressibility, and not the vanishing compressibility we observe. It is interesting that the integrity of the $\nu = 1/2$ state is more sensitive to disorder than $\nu = 1/3$ and $\nu = 2/3$, especially at the smaller values of $d/a_M$. We see this same behavior in the experiment when we compare the two measurements at 10 K in Fig. 2.

**Device entropy.** The energy density $u$ in the above treatment has to be replaced by the free energy density $f$ at finite temperatures. Eqn. (6) now becomes

$$V_t = \frac{2}{e}\left(\frac{\partial f}{\partial n}\right)_T. \tag{22}$$

We can therefore obtain $\left(\frac{\partial V_t}{\partial T}\right)_\nu = -\frac{2}{e}\left(\frac{\partial S}{\partial \nu}\right)_T$ (known as the Maxwell's relation) and the entropy per moiré unit cell [41,42]

$$S = -\frac{e}{2}\int d\nu \left(\frac{\partial V_t}{\partial T}\right)_\nu \tag{23}$$

from the data in Fig. 3. Note that because the MoSe$_2$/WS$_2$ band gap is temperature dependent, the $V_t$ threshold for the onset of charging shifts with temperature. Over a certain temperature range, this shift only adds a constant to $\left(\frac{\partial V_t}{\partial T}\right)_\nu$ but does not contribute to the electronic entropy. We have therefore subtracted a temperature-independent constant from $\left(\frac{\partial V_t}{\partial T}\right)_\nu$ to obtain a filling-dependent entropy such that $S = 0$ at $\nu = 0$ and $\nu = 1$. This is justified because of the negligible electronic DOS of the metal gate compared to MoSe$_2$/WS$_2$ and because of the large energy gap at these filling factors ($\nu = 0$ and $\nu = 1$ correspond to the MoSe$_2$/WS$_2$ band gap and the Mott gap, respectively), where the electronic entropy of MoSe$_2$/WS$_2$ is expected to vanish at temperatures much below the gap size.

**Figures**

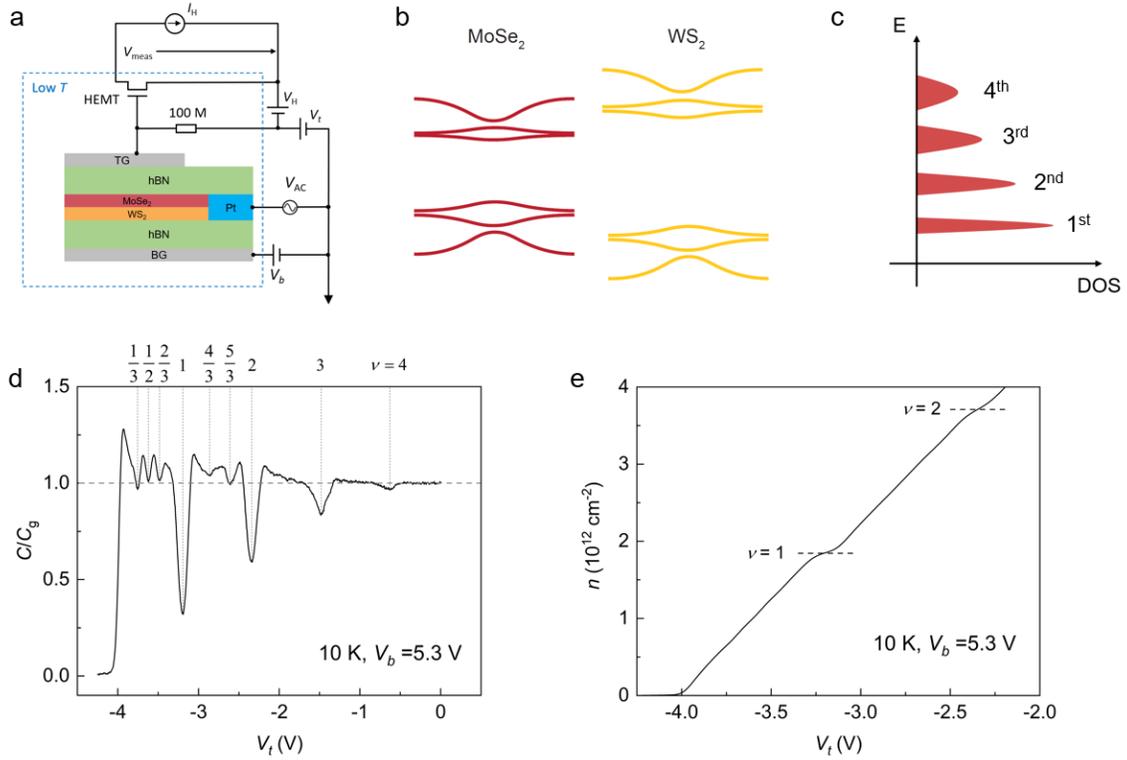

**Figure 1 | Charge-order-enhanced capacitance in MoSe$_2$/WS$_2$ moiré superlattices. a,** Schematics of a dual-gate device structure and electrical connections for capacitance measurements. TG and BG denote the top and back gate, respectively. **b,** Schematic moiré miniband structure of MoSe$_2$/WS$_2$, which has a type I band alignment. **c,** Schematic illustration of density of states (DOS) versus energy for the first four Hubbard bands. The bandwidth increases and the DOS decreases with energy. **d,** Differential top gate capacitance as a function of top gate voltage at 10 K for device 1 ($d/a_M \approx 1$). The back gate voltage is fixed at 5.3 V. The filling factors for discernable incompressible states are labeled on the top axis. **e,** Charge carrier density as a function of top gate voltage obtained from integrating **d**. Density plateaus at incompressible state $\nu = 1$ and 2 are labeled.

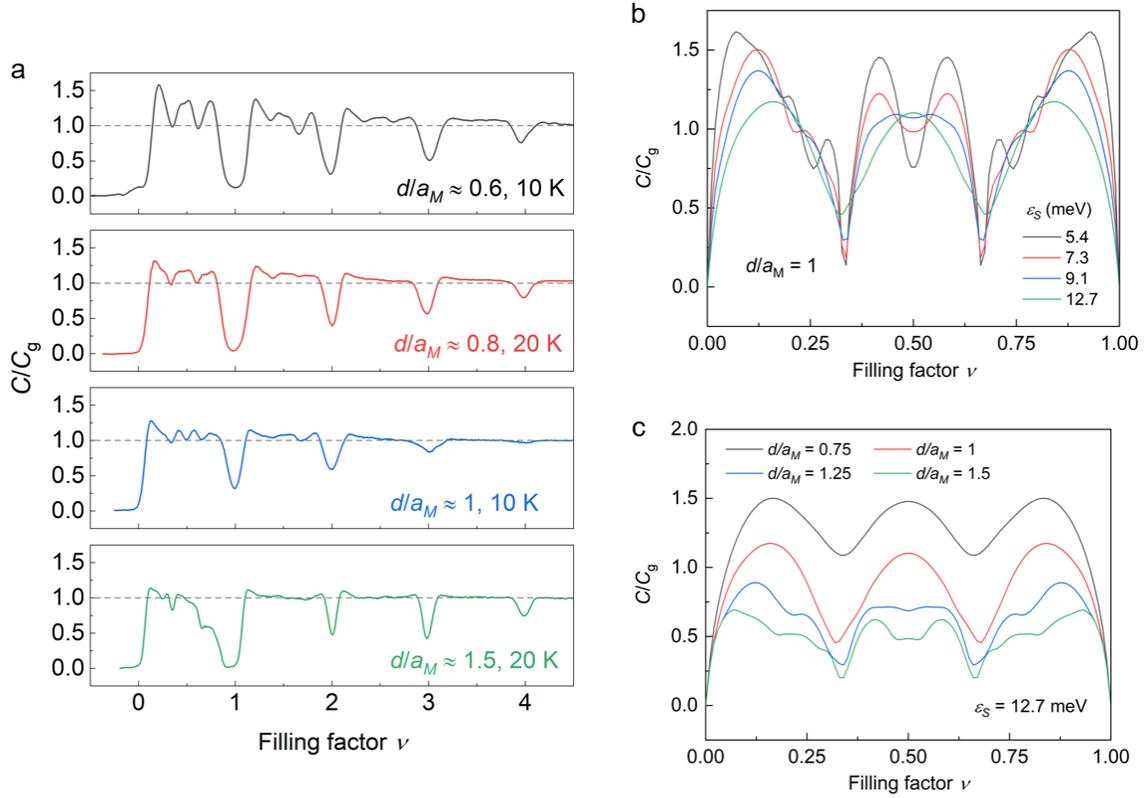

**Figure 2 | Dependence on sample-gate separation. a,** Experimental top gate capacitance as a function of filling factor for devices with $d/a_M \approx 0.6, 0.8, 1.0$ and $1.5$ (from top to bottom) at the lowest temperature (10 K or 20 K) allowed by the sample/contact resistance. The $d/a_M \approx 1$ device has the smallest sample/contact resistance among all devices (Extended Data Fig. 3). The capacitance enhancement in the compressible regions decreases with increasing $d/a_M$. **b, c,** Calculated top gate capacitance as a function of filling factor up to $\nu = 1$ for different disorder strengths for $d/a_M = 1$ (**b**) and for different $d/a_M$ with a fixed disorder strength $\varepsilon_S = 12.7$ meV (**c**).

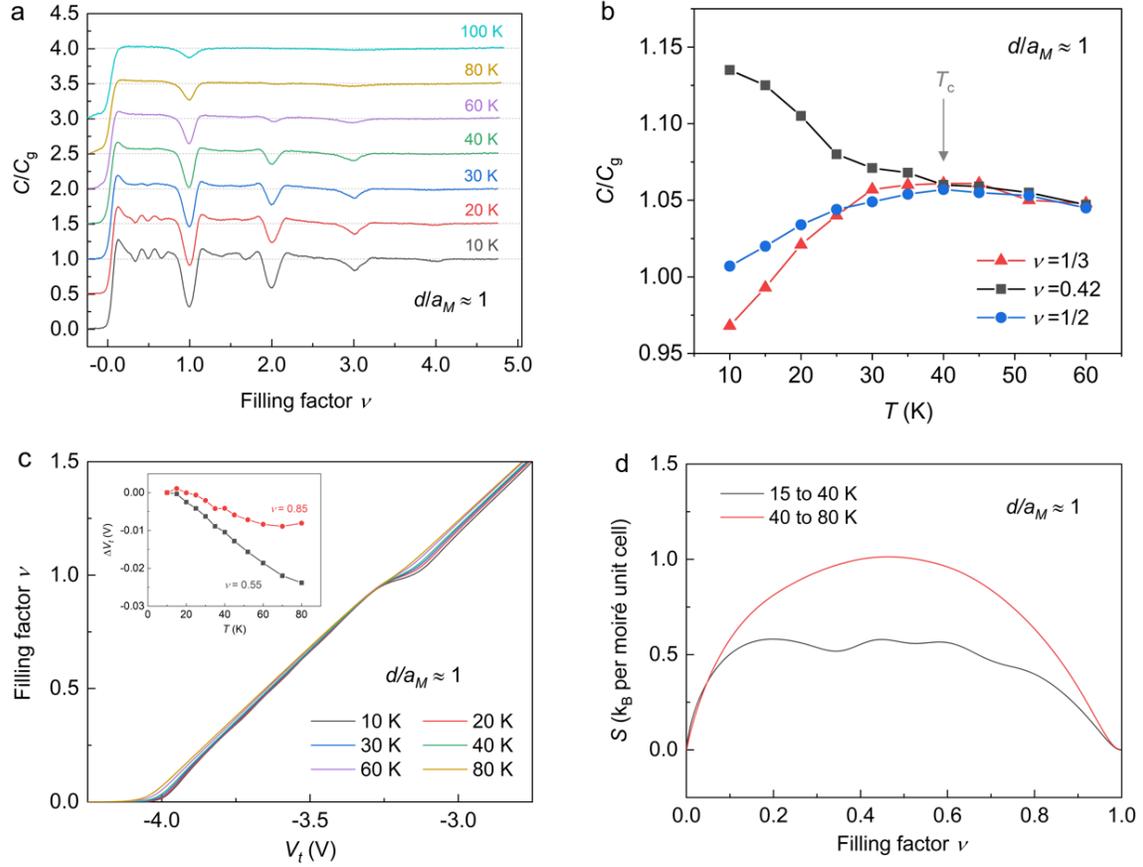

**Figure 3 | Temperature dependence and device entropy. a,** Experimental top gate capacitance as a function of filling factor at different sample temperatures for device 1 ($d/a_M \approx 1$). The charge-ordered states melt and the large capacitance enhancement disappears at high temperatures. **b,** Top gate capacitance at selected filling factors as a function of temperature. The capacitance in the incompressible region ($\nu = 1/3$ and $1/2$) decreases rapidly below the melting temperature $T_C$ of the charge-ordered state, whereas the capacitance in the compressible regions ($\nu \approx 0.42$) increases. **c,** Filling factor versus top gate voltage at different temperatures obtained from integrating the data in **a**. The temperature dependence of the top gate voltage at constant filling factors is shown in the inset. **d,** Filling factor dependence of the device entropy (per moiré unit cell) extracted from two temperature windows: 15-40 K (black) and 40-80 K (red).

**Table 1 | Transition temperature $T_C$ and thermodynamic gap $\Delta\mu$ of the incompressible states in MoSe$_2$/WS$_2$ device 1 ($d/a_M \approx 1$).** The thermodynamic gap of the integer-filling states are extrapolated from the temperature dependence of $\Delta\mu$ down to 0 K. The thermodynamic gap of the sample at fractional fillings cannot be determined because of sample-gate coupling. The uncertainty for $T_C$ and $\Delta\mu$ is estimated to be $\pm 5$ K and $\pm 2$ meV, respectively.

|  | $\nu$ | 1/3 | 1/2 | 2/3 | 1 | 4/3 | 5/3 | 2 | 3 | 4 |
|---|---|---|---|---|---|---|---|---|---|---|
| Device 1 | $T_C$ (K) | 40 | 40 | 40 | 160 | 30 | 30 | 75 | 80 | 25 |
| ($d/a_M \approx 1$) | $\Delta\mu$ (meV) | ~ | ~ | ~ | 58.5 | ~ | ~ | 36.3 | 18.1 | 4.8 |

**Extended Data Figures and Table**

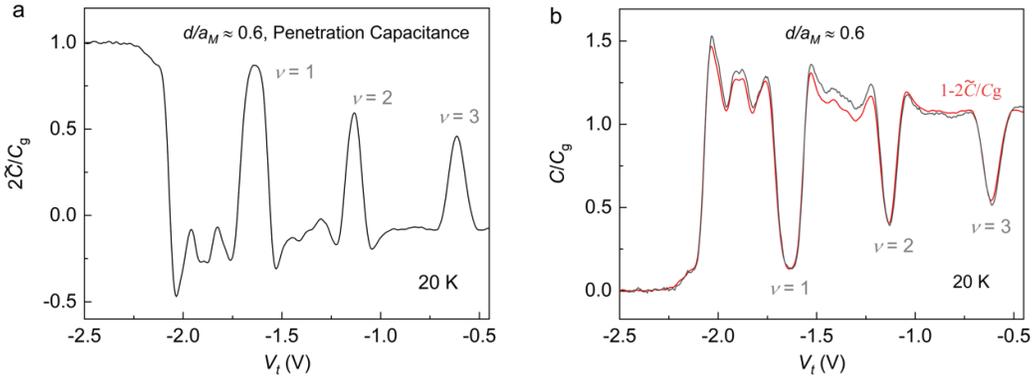

**Extended Data Figure 1 | Penetration capacitance.** Penetration capacitance $2\tilde{C}/C_g$ (**a**) and top gate capacitance $C/C_g$ (**b**) as a function of top gate voltage at 20 K ($d/a_M \approx 0.6$). We also show $1 - 2\tilde{C}/C_g$ (obtained from **a**) in **b** for comparison. Good agreement between penetration and top gate capacitance measurements is observed.

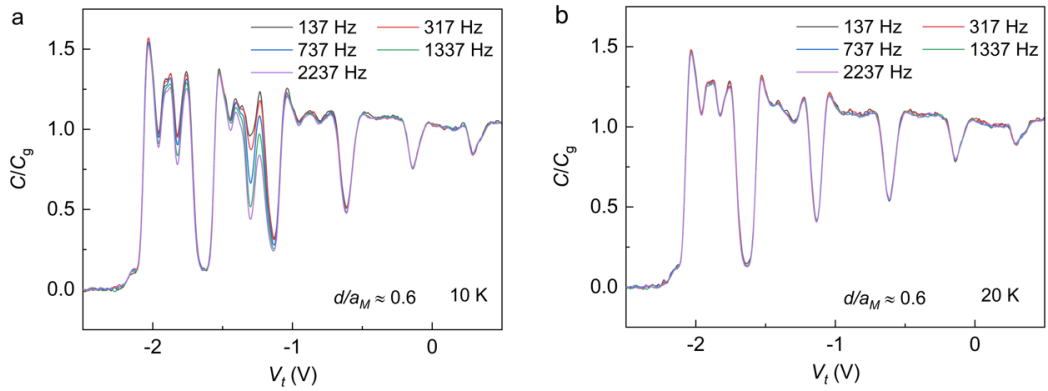

**Extended Data Figure 2 | Frequency dependent capacitance ($d/a_M \approx 0.6$). a,** Top gate capacitance $C/C_g$ as a function of top gate voltage at 10 K (**a**) and at 20 K (**b**) under different excitation frequencies (10 mV modulation amplitude). A decrease in capacitance with increasing frequency is observed at 10 K, signifying the growing importance of the resistive component in the AC measurement. Meanwhile, no obvious frequency dependence is seen at 20 K, a temperature at which the sample/contact resistance becomes substantially smaller. The capacitance measurement is therefore in the low-frequency limit and reliable for most of the filling factors except $\nu = 1$ and $\nu = 2$.

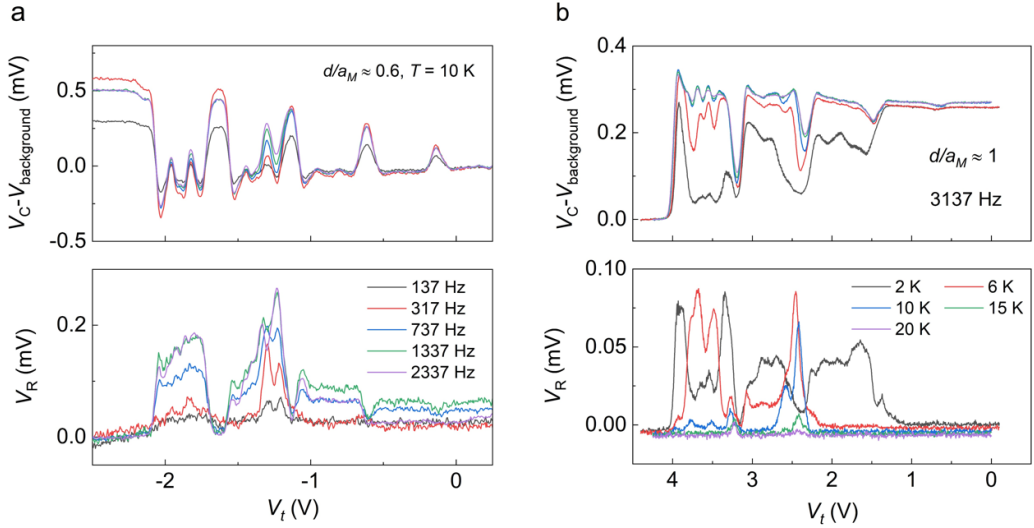

**Extended Data Figure 3 | Frequency and temperature dependent capacitive and resistive channels. a,** Dependence of the measured output voltage in the capacitive channel (top) and the resistive channel (bottom) at different excitation frequencies ($d/a_M \approx 0.6$ and 10 K). We have subtracted a constant background voltage for the capacitive channel for better comparison. We can see that the resistive signal grows with frequency for most of the filling factors except $\nu = 1$ and $\nu = 2$, where a slight decrease in the resistive signal with frequency is observed. We can therefore conclude that the AC capacitance measurement at 10 K is reliable for most of the filling factors at 137 Hz (for this device, device 4). For $\nu = 1$ and $\nu = 2$, however, the frequency dependence suggests that the AC measurement is in the high-frequency limit so that the measured capacitance dips at $\nu = 1$ and $\nu = 2$ only reflects the large sample/contact resistance. **b,** Dependence of the measured output voltage in the capacitive channel (top) and the resistive channel (bottom) at different sample temperatures ($d/a_M \approx 1$ and excitation frequency at 3137 Hz). As the resistive signal grows with decreasing sample temperature, the capacitive signal decreases accordingly. The results (except at 2 K) are largely consistent with measurements in the low-frequency limit. At and above 10 K, the resistive signal is negligible compared to the capacitive signal. Furthermore, small resistive peaks at $\nu = 1$ and $\nu = 2$ that grow with decreasing temperature are observed. All of these show that the measurements are in the low-frequency limit for all filling factors in this device (Device 1) and capacitance measurements at and above 10 K are reliable.

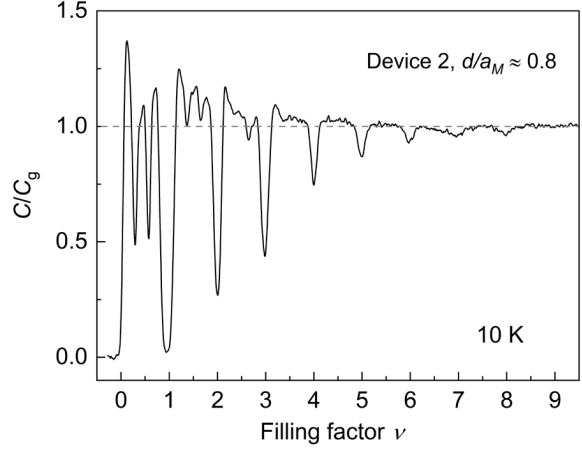

**Extended Data Figure 4 | Filling factor dependence of the top gate capacitance (device 2) showing incompressible states up to $\nu = 8$.**

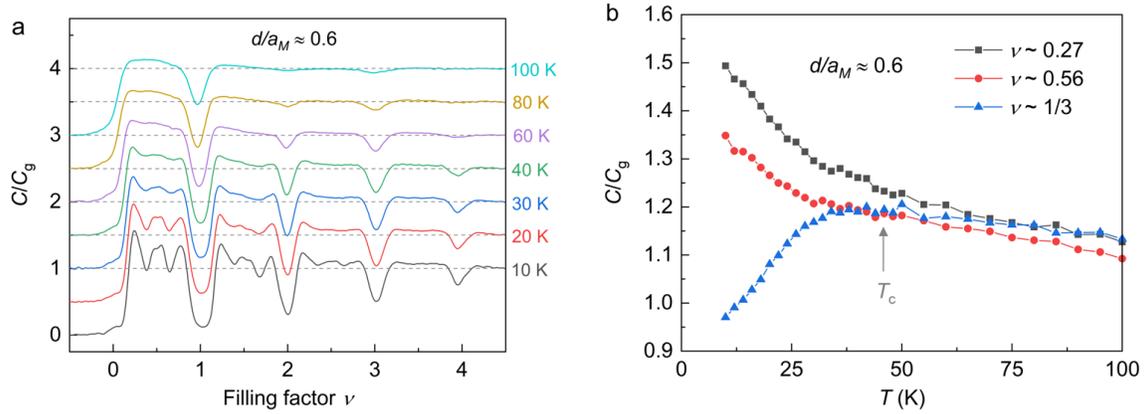

**Extended Data Figure 5 | Temperature dependence for another device ($d/a_M \approx 0.6$). a,** Experimental top gate capacitance as a function of the filling factor at different sample temperatures. **b,** Top gate capacitance at selected filling factors as a function of temperature. While the capacitance at the compressible region ($\nu \approx 0.27$ and $0.56$) increases quickly below $T_C$ of the charge-ordered states, that at the incompressible regions ($\nu \approx 1/3$) decreases. The results are similar to those in Fig. 3 in the main text.

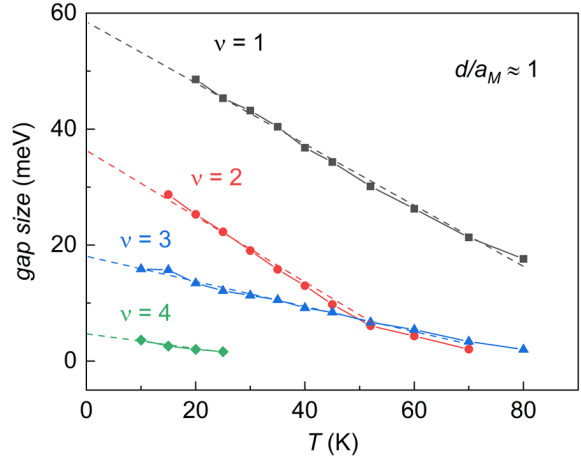

**Extended Data Figure 6 | Temperature dependence of Δμ (device 1).** Extracted thermodynamic gap size $\Delta\mu$ as a function of temperature for $\nu = 1 - 4$. A linear dependence is seen at low temperatures. Extrapolation by fitting to a linear curve allows the extraction of the zero-temperature gap size tabulated in Table 1.

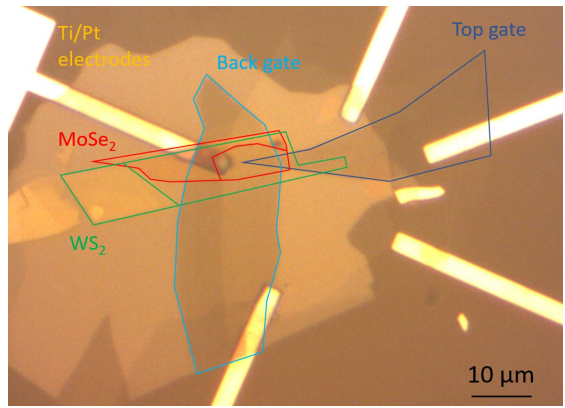

**Extended Data Figure 7 | Optical image of a typical device.**

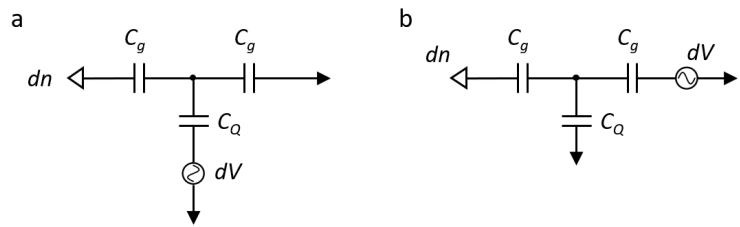

**Extended Data Figure 8 | Lumped circuit model for top gate capacitance measurement (a) and penetration capacitance measurement (b).**

**Extended Data Table 1 | Calculated chemical potentials (in units of $\epsilon_a$) labeled by filling, $\mu_\nu$, at different filling factors and $d/a_M$.**

| $\nu$ | $d/a_M = 0.75$ $\mu_\nu$ | $d/a_M = 1.00$ $\mu_\nu$ | $d/a_M = 1.25$ $\mu_\nu$ | $d/a_M = 1.50$ $\mu_\nu$ |
|---|---|---|---|---|
| 1/7 | 0.017 | 0.059 | 0.126 | 0.211 |
| 1/4 | 0.076 | 0.198 | 0.357 | 0.538 |
| 1/3 | 0.146 | 0.339 | 0.573 | 0.828 |
| 2/5 | 0.269 | 0.525 | 0.821 | 1.137 |
| 1/2 | 0.409 | 0.754 | 1.140 | 1.547 |
| 3/5 | 0.538 | 0.973 | 1.449 | 1.947 |
| 2/3 | 0.611 | 1.103 | 1.639 | 2.197 |
| 3/4 | 0.743 | 1.311 | 1.922 | 2.556 |
| 6/7 | 0.900 | 1.566 | 2.275 | 3.006 |
| 1 | 1.076 | 1.867 | 2.705 | 3.566 |